\documentstyle[12pt]{article}

\newtheorem{theorem}{Theorem}
\newtheorem{lemma}{Lemma}

\begin{document}

\newcommand{\lap}{\bigtriangleup}
\def\be{\begin{equation}}
\def\ee{\end{equation}}
\def\bea{\begin{eqnarray}}
\def\eea{\end{eqnarray}}
\def\beas{\begin{eqnarray*}}
\def\eeas{\end{eqnarray*}}
\def\n#1{\vert #1 \vert}
\def\nn#1{{\Vert #1 \Vert}}
\def\R{{\rm I\kern-.1567em R}}
\def\N{{\rm I\kern-.1567em N}}
 
\def\supp{\mbox{\rm supp}\,}
\def\suppi{\mbox{\scriptsize supp}\,}
\def\ekin{E_{\rm kin}}
\def\epot{E_{\rm pot}}

\def\D{{\cal D}}
\def\C{{\cal C}}
\def\X{{\cal X}}
\def\F{{\cal F}}
\def\P{{\cal P}}
\def\M{{\cal M}}
\def\prfe{\hspace*{\fill} $\Box$

\smallskip \noindent}

\title{Existence and stability of Camm type 
       steady states in galactic dynamics}
\author{ Yan Guo\\
         Lefschetz Center for Dynamical Systems \\
         Division of Applied Mathematics \\
         Brown University, Providence, RI 02912 \\
         and \\
         Gerhard Rein\\
         Mathematisches Institut 
         der Universit\"at M\"unchen\\
         Theresienstr. 39\\
         80333 M\"unchen, Germany}
\date{}
\maketitle

\begin{abstract}
We prove the existence and nonlinear stability of Camm type
steady states of 
the Vlasov-Poisson system in the gravitational case. 
The paper demonstrates the effectiveness of an approach to the
existence and stability problem for steady states,
which was used in previous work by the authors:
The steady states are obtained as minimizers of an energy-Casimir 
functional, and from this fact their dynamical stability 
is deduced. 
\end{abstract}

\section{Introduction}
\setcounter{equation}{0}

In astrophysics the Vlasov-Poisson system 
\be \label{vlasov}
\partial_t f + v \cdot \nabla _x f - \nabla _x U \cdot 
\nabla _v f = 0,
\ee
\be \label{poisson} 
\lap U = 4 \pi\, \rho, 
\ee
\be \label{rho}
\rho(t,x)= \int f(t,x,v)dv 
\ee
is used to model the time evolution of a large ensemble of
``particles'' (stars)  which interact only by the gravitational
field which 
they create collectively. Examples of such ensembles are
galaxies or globular clusters.
Here $f = f(t,x,v)\geq 0$ denotes the density of the particles 
in phase space, $t \in \R$ denotes time, $x, v \in \R^3$ denote 
position and velocity respectively, $\rho$ is the spatial mass 
density, and $U$ the gravitational potential. 
The model does not include relativistic effects---including these
would lead to the
Vlasov-Einstein system---or collisions among the particles---these
are assumed to be sufficiently rare to
be neglected.  

In the present paper we are interested in the existence and
stability of steady states of this system,
and we pursue an approach which has recently been used 
to construct stable steady states of polytropic type
and generalizations of these, cf.\ \cite{G,GR}. For polytropic steady 
states the phase space density is of the form
\be \label{poly}
f(x,v) = (E_0 - E)_+^k L^l.
\ee
Here $(\cdot)_+$ denotes the positive part, $E_0 \in \R$ is a constant, 
\be \label{parten}
E=\frac{1}{2}|v|^2 + U (x) 
\ee
denotes the particle energy which is conserved along characteristics of the Vlasov equation (\ref{vlasov}) if $U$ is time-independent, and 
\be \label{angmom}
L=|x \times v|^2 = |v|^2|x|^2-(x \cdot v)^2,
\ee
denotes the modulus of angular momentum squared which is conserved if 
$U$ is spherically symmetric.
Upon substitution of the ansatz (\ref{poly}) into (\ref{rho})
the Vlasov-Poisson system
is reduced to the---then semilinear---Poisson equation. This approach  
was followed in \cite{BFH}, where it was shown that solutions of the
semilinear Poisson equation exist and lead to steady states with finite 
mass and compact support, provided 
$k>-1,\ l>-1,\ k+l+3/2 \geq 0,\ k< 3l +7/2$.
The question whether the resulting steady states are stable is not addressed
by this approach. 
In the present paper we construct steady states as minimizers of
an appropriately defined energy-Casimir functional.
This has several advantages: The fact that the resulting steady states
have finite mass is built into the definition of the set over which one
minimizes the energy-Casimir functional, and the compact support 
property is an integral part of the minimization approach as well.
Next, the appoach is more flexible in the sense that one does not
need an ansatz exactly of the form (\ref{poly}), but only certain growth and
scaling assumptions. Finally and most importantly, resulting steady
states are stable in a well defined sense. In
\cite{G} steady states of the form (\ref{poly}) with $0<k<l+3/2$
were considered, and in \cite{GR} this was extended to include
steady states of the form $f(x,v) = \phi (E_0-E,L)$ where $\phi$
is characterized by certain growth conditions. An extension of the
polytropic ansatz is 
\be \label{camm}
f(x,v)=(E_0 - E - \gamma L)_+^k L^l,
\ee
which is due to Camm \cite{C}; here $\gamma \geq 0$ is an additional parameter.
We will show that the energy-Casimir technique applies
for $k$ and $l$ as above and $\gamma$ small and yields steady states with finite mass and compact support which are nonlinearly stable.
At the same time the method will allow a more general  
dependence on $E_0 - E - \gamma L$. 
In the case $\gamma =0$ the present paper includes steady states
which were not covered in \cite{G,GR}. As indicated by \cite[5.8]{BFH},
a smallness assumption on $\gamma$ is necessary to obtain steady states 
with compact support. 
 
The paper proceeds as follows. In the next section
we  introduce the energy-Casimir functional $\D$ and 
prove some preliminary results, in
particular a lower bound for $\D$ on an appropriate set $\F_M$
of test functions  with prescribed mass $M$. 
The crucial part is to show that along
a minimizing sequence mass cannot escape to infinity. 
This is done in Section 3, using the scaling properties of $\D$
and a careful estimate of the contribution of
the part of the matter distribution inside and the part outside a given
ball in space to the energy-Casimir functional.
For this splitting estimate we require spherical symmetry 
of the functions in $\F_M$.
In Section 4 we show that a minimizer exists and that
any minimizer is a steady state, the latter fact being essentially the
Euler-Lagrange identity for our variational problem.
In the last section we discuss the stability properties of the
resulting steady states. 
 
We conclude this introduction with some further references to the 
literature. The existence of global classical solutions 
to the initial value problem for the Vlasov-Poisson system has been 
shown in \cite{P} as well as in \cite{LP,S}.  
In the monograph \cite{FP} one can find many references 
to discussions of the stability problem in the astrophysics
literature. As far as
mathematically rigorous results are concerned, we mention
\cite{Wo}, where the stability of the polytropes is investigated
using a variational approach for a reduced energy-Casimir
functional defined on the space of mass functions
$m(r)=4 \pi \int_0^r s^2 \rho(s)\, ds$, and an investigation
of linearized stability in \cite{BMR}. For the plasma physics case,
where the sign in the Poisson equation (\ref{poisson}) is
reversed, the stability problem is much easier and better
understood. We refer to \cite{BRV,GS1,GS2,R2}. The present approach
was also used in \cite{R} to show the existence and stability
of extremely flattened steady states which in particular are no
longer spherically symmetric. Finally, a very general condition
which guarantees finite mass and compact support of steady
states, but not their stability, is established in \cite{RR}.

\section{Preliminaries; a lower bound for $\D$}
\setcounter{equation}{0}

For a measurable, spherically symmetric
function $f=f(x,v),\ x \in \R^3$, and $r=\n{x}$ we define
\[
\rho_f (x):= \int f(x,v)\, dv,\ m_f (x) := \int_{\n{y} \leq r}
\rho_f(y)\, dy,
\]
and
\[  
\nabla U_f (x) :=  U_f'(r) \frac{x}{r}
:=\frac{m_f(r)}{r^2}  \frac{x}{r},\ 
U_f(r) := - \int_r^\infty U_f'(s)\, ds .  
\]
Here spherical symmetry means that
\[
f(Ax,Av)=f(x,v),\ x,v \in \R^3,\ A \in \mbox{\rm SO}(3);
\]
the symmetry is of course only relevant for the definition of the potential.
We shall also use the notation $m_\rho$ and $U_\rho$
if $\rho$ is not necessarily induced by some function $f(x,v)$.
Spherically symmetric functions of $x$ will be identified with the 
corresponding functions of $r=\n{x}$.
Next we define
\beas
\ekin (f)
&:=&
\frac{1}{2} \int\!\!\int \n{v}^2 f(x,v)\,dv\,dx,\\
\epot (f)
&:=&
- \frac{1}{8\pi} \int |\nabla U_f (x)|^2 dx,\\
\C(f)
&:=&
\int\!\!\int Q(L^{-l}f(x,v))L^l\,dv\,dx,\\
\P(f)
&:=&
\gamma\int L f(x,v)\,dv\,dx + \C(f) + \ekin(f)  ,\\
\D(f)
&:=&
\P(f)  + \epot (f),
\eeas
where $l>-1$,
$Q$ is a given function satisfying certain assumptions specified 
below, and $\gamma \geq 0$. Note that $\P$ is the positive part of the energy-Casimir functional $\D$. As to the existence of the potential energy 
part we refer to Lemma~\ref{uest} below.
We will minimize $\D$ over the set
\bea \label{spacedef}
\F_M := \Bigl\{ f \in L^1(\R^6) 
&\mid&
f \geq 0,\ \int\!\!\int f dv\,dx = M, \nonumber \\
&& 
\P(f) < \infty,\ \mbox{and}\ f\ \mbox{is spherically symmetric}\Bigr\},
\eea
where $M>0$ is prescribed. The function $Q$ which determines the Casimir functional has to satisfy
the following  

\smallskip
\noindent {\bf Assumptions on $Q$}: 
$Q \in C^1 ([0,\infty[)\cap 
C^2 (]0,\infty[)$, $Q \geq 0$, and there exist
constants $C_1,\,C_2 >0$, $F_0 >0$, and 
$0 < k_1,\,k_2,\,k_3 < l + 3/2$ such that:
\begin{itemize}
\item[(Q1)]
$Q(f) \geq C_1 f^{1+1/{k_1}},\ f \geq 0$; if $l=0$ this is required 
for $f \geq F_0$ only.
\item[(Q2)]
$Q(f) \leq C_2 f^{1+1/{k_2}},\ 0 \leq f \leq F_0$.
\item[(Q3)]
$Q(\lambda f) \geq \lambda^{1+1/{k_3}} Q(f),\ f \geq 0,\ 
0 \leq \lambda \leq 1$. 
\item[(Q4)]
$ Q''(f) > 0,\ f > 0$, and $ Q'(0) = 0$.
\end{itemize}

\noindent
{\bf Remark:} The above assumptions imply that
$Q'$ is strictly increasing with range $[0,\infty[$.
On their support
the steady states obtained later will be of the form 
\[
f_0 (x,v) = (Q')^{-1}(E_0 -E - \gamma L) L^l 
\]
with some $E_0<0$ and $E$ and $L$ as defined in 
(\ref{parten}) and (\ref{angmom}) respectively.
A typical example of a function $Q$ satisfying the assumptions
would be
\be \label{nocamm}
Q(f) = c_1 f^{1+1/{k_1}} +  c_2 f^{1+1/{k_2}}
\ee
with $0 < k_1,\,k_2 < l + 3/2$ and $c_1 >0$, $c_2 \geq 0$.
For $c_2 = 0$ this leads to a steady state of the form (\ref{camm}),
but this is not so if $c_2>0$. 

\smallskip

The aim of the present section is to establish a lower bound
for $\D$ of a form that will imply the boundedness of $\P$
along any minimizing sequence. On the way we will establish several estimates
for $\rho_f$ and $U_f$ induced by an element $f \in \F_M$.

\begin{lemma} \label{rhoest}
\begin{itemize}
\item[{\rm (a)}]
There exists a constant $C >0$ such that
\[
\int\!\!\int f^{1+1/k_1} L^{-l/k_1} dv\,dx \leq C (1 + \P(f)),\ f \in \F_M .
\]
\item[{\rm (b)}]
Let $n_1 := k_1 +l +3/2$. Then there exists a constant
$C>0$ such that
\[
\int \rho_f^{1+1/n_1} \n{x}^{-2l/n_1} dx \le C 
\left( 1 + \P (f) \right),\ f \in \F_M.
\]
\end{itemize}
\end{lemma}

\noindent
{\bf Proof}.  If (Q1) holds with $F_0=0$ the estimate in (a)
is obvious. If $l=0$ and $F_0>0$ one can split the integral
according to $f \leq F_0$ and $f \geq F_0$ and use $M$ to bound
the first part and (Q1) to bound the second part. As to (b),
we have for any $R>0$ and $x \in \R^3$,
\beas
\rho_f (x) 
&=&
\int_{\n{v} \leq R} f (x,v)\,dv + \int_{\n{v} \geq R} f (x,v)\,dv\\
&\leq&
\biggl(\int_{\n{v} \leq R} L^l dv\biggr)^{1/(1+k_1)}\left(\int f^{1+1/k_1} L^{-l/k_1} dv \right)^{k_1/(k_1+1)}
+ \frac{1}{R^2} \int v^2 f\, dv\\
&=&
C \n{x}^{2l/(k_1 + 1)}R^{(2l+3)/(k_1+1)}
\left(\int f^{1+1/k_1} L^{-l/k_1} dv \right)^{k_1/(k_1+1)}
 +  \frac{1}{R^2} \int v^2 f\, dv \\
&\leq&
C \n{x}^{2l/(k_1 + l + 5/2)} \left(\int f^{1+1/k_1} L^{-l/k_1} dv
+ \int v^2 f\, dv \right)^{n_1/(n_1+1)}
\eeas
by H\"older's inequality and optimization in $R$. 
Taking both sides of
the inequality to the power $1+1/n_1$,
dividing by $\n{x}^{2l/n_1}$, and integrating with respect to $x$
yields the assertion.
\prfe

Motivated by Lemma~\ref{rhoest} we define 
\beas
L^{k_1,l}(\R^6) :=
\Bigl\{f : \R^6 \to \R 
&\mid&
 f \ \mbox{measurable, spherically symmetric, and}\\
&&
\int\!\!\int f^{1+1/k_1} L^{-l/k_1} dv\,dx < \infty \Bigr\}
\eeas
equipped with the norm
\[
\nn{f}_{k_1,l} := \left(\int\!\!\int f^{1+1/k_1} L^{-l/k_1} dv\,dx   \right)^{k_1/(k_1+1)},
\]
and
\beas
L^{n_1,l}(\R^3) :=
\Bigl\{ \rho : \R^3 \to \R 
&\mid&
 \rho \ \mbox{measurable, spherically symmetric, and}\\
&&
\int \rho^{1+1/n_1} \n{x}^{-2l/n_1} dx < \infty \Bigr\}
\eeas
equipped with the norm
\[
\nn{\rho}_{n_1,l} := \left(\int \rho^{1+1/n_1} \n{x}^{-2l/n_1} dx \right)^{n_1/(n_1+1)} .
\]
Both spaces are reflexive Banach spaces. 

\begin{lemma} \label{uest}
\begin{itemize}
\item[{\rm (a)}]
There exist constants $C>0$ and $q>0$ such that
for $\rho \in L^{n_1,l}(\R^3)$  
with $\int |\rho| = M$ we have 
\beas
\int |\nabla U_\rho|^2 dx
&\leq&
4 \pi \int_0^R \frac{m^2_\rho (r)}{r^2} \, dr  
+ \frac{4 \pi M^2}{R}\\ 
&\leq&
C R^q \left(1 + \nn{\rho}_{n_1,l}^{1+1/n_1} \right) + \frac{4 \pi M^2}{R},\
R>0.
\eeas
\item[{\rm (b)}]
For every $R>0$ the mapping 
\[
T : L^{n_1,l}(\R^3) \ni \rho \mapsto \frac{m_\rho}{r}\Bigr|_{[0,R]}
\in L^2([0,R])
\]
is compact.
\item[{\rm (c)}]
For $\rho_1,\,\rho_2 \in L^{n_1,l}(\R^3) \cap L^1(\R^3)$ 
the following identity holds:
\[
\int \nabla U_{\rho_1} \cdot \nabla U_{\rho_2} dx = 
- 4 \pi\,\int U_{\rho_1} \rho_2 dx.
\]
\end{itemize} 
\end{lemma}

\noindent
{\bf Proof.}
H\"older's inequality implies that for any $\rho \in L^{n_1,l}(\R^3)$, 
\be \label{mest}
|m_\rho (r)| \leq (4 \pi)^{1/(1+n_1)}
\nn{\rho}_{n_1,l} r^{(2l+3)/(n_1+1)},\ r \geq 0,
\ee
and thus
\be \label{tbound}
\int_0^R \frac{m^2_\rho (r)}{r^2} \, dr
\leq C \nn{\rho}_{n_1,l}^2 R^{(4l+5-n_1)/(n_1+1)} .
\ee
The first estimate for $\nabla U_\rho$ in (a) follows from
spherical symmetry and the fact that $|m_\rho|\leq M$.
For $n_1\leq 1$ the second estimate immediately follows
from (\ref{tbound}).
For $n_1 > 1$ we use  $|m_\rho|\leq M$ and (\ref{mest}) to obtain
\[
\int_0^R \frac{m^2_\rho (r)}{r^2} \, dr
\leq M^{1-1/n_1} \int_0^R \frac{|m_\rho|^{1+1/n_1} (r)}{r^2} \, dr
\leq
C \nn{\rho}_{n_1,l}^{1+1/n_1} R^{(2l+3-n_1)/n_1} .
\]
Since in both cases
the power of $R$ is positive, this proves (a). 
As to (b), we first observe that the operator $T$ is bounded
by (\ref{tbound}).
To show its compactness we take a bounded set $S \subset L^{n_1,l}(\R^3)$ and
apply the Frech\'{e}t-Kolmogorov criterion to the set $TS$. We redefine
$T\rho := \frac{m_\rho}{r} 1_{[0,R]} \in L^2(\R)$ where $1_{[0,R]}$
is the characteristic function of the interval $[0,R]$.
The crucial part is so show that 
\[
\nn{ (T\rho)_h - T\rho}_2 \to 0,\ h \to 0
\]
uniformly in $\rho \in S$, where $(T\rho)_h = (T\rho)(\cdot +h)$.
For $h>0$ this follows from the estimate
\beas
\left|\left|\left(\frac{m_\rho}{r} 1_{[0,R]}\right)_h - \frac{m_\rho}{r} 1_{[0,R]}\right|\right|_2^2
&\leq&
2 \int_0^h  \frac{m^2_\rho}{r^2}\, dr + \int_{R-h}^R \frac{m^2_\rho}{r^2}\, dr\\
&&
{}+ \int_h^{R-h} m^2_\rho (r) \left|\frac{1}{r+h} - \frac{1}{r} \right|^2 dr\\
&&
{}+ \int_h^{R-h} \frac{1}{(r+h)^2}\n{m_\rho(r+h) - m_\rho (r)}^2 dr.\\
\eeas
For the first three terms one uses the estimate (\ref{mest}).
By H\"older's inequality
\[
\n{m_\rho (r+h) - m_\rho (r)} \leq C \nn{\rho}_{n_1,l}
\left((r+h)^{2l+3} - r^{2l+3}\right)^{1/(n_1+1)} ,
\]
and together with Lebesgue's dominated convergence theorem this yields
the convergence of the last term. Obviously, each term converges
uniformly in $\rho \in S$, and the case $h<0$ is completely analogous.
As to part (c), we have
\beas
\int \nabla U_{\rho_1} \cdot \nabla U_{\rho_2} dx
&=&
4 \pi\, \int_0^\infty U_{\rho_1}'(r)\, m_{\rho_2} (r)\, dr\\
&=&
4 \pi\, U_{\rho_1}(r)\, m_{\rho_2} (r) \biggl|_{r=0}^{r=\infty} -
(4 \pi)^2 \int_0^\infty U_{\rho_1}(r)\, r^2 \rho_2 (r)\, dr\\
&=&
{} - 4 \pi\,\int U_{\rho_1} \rho_2 dx .
\eeas
Here the boundary term at infinity vanishes since $|U_{\rho_1} (r)| \leq \nn{\rho_1}_1/r$
and $|m_{\rho_2} (r)| \leq \nn{\rho_2}_1$, and the boundary term at zero 
vanishes by (\ref{mest}). \prfe

\begin{lemma} \label{lower}
There exists a constant $C>0$ such that
\[
\D (f) \geq {1\over 2} \P(f) - C,\ f \in \F_M,
\]
in particular,
\[
\D_M := \inf\, \{\D(f) \mid f \in \F_M\} > - \infty .
\]
\end{lemma}

\noindent
{\bf Proof.} 
Using the previous two lemmas we have 
\beas
\D(f)
&\geq&
\P(f) - C  R^q (1 + \nn{\rho_f}_{n_1,l}^{1+1/n_1}) - \frac{4\pi M^2}{R}\\
&\geq&
\P(f) (1-CR^q) - C R^q -  \frac{4\pi M^2}{R},
\eeas
where $C>0$ is some constant which does not depend on $R>0$.
The assertion follows by a suitable choice of $R$. \prfe

\section{Scaling and splitting}
\setcounter{equation}{0}

The behaviour of $\D$ and $M$ under scaling transformations can be
used to show that $\D_M$ is negative
for $\gamma$ small and to
relate the $\D_M$'s for different values of $M$:

\begin{lemma} \label{scaling}
\begin{itemize}
\item[{\rm (a)}]
Let $M>0$. Then $-\infty < \D_M  < 0$
for $\gamma\geq 0$ sufficiently small.
\item[{\rm (b)}]
There exists $\alpha >0$ such that for all $\gamma \geq 0$
and $ 0< M_1 \leq M_2$,
\[
\D_{M_1}\ge \left( {{M_1}\over{M_2}}\right )^{1+\alpha}\D_{M_2}.
\]
\end{itemize}
\end{lemma}

\noindent
{\bf Proof.} 
Given any function $f$, we define a rescaled function 
$\bar f(x,v)=af(bx,cv)$, where $a,\,b,\,c >0$. Then
\be \label{mscale}
\int\!\!\int \bar f\,dv\,dx = ab^{-3}c^{-3}\int\!\!\int f\,dv\,dx
\ee
and
\bea
\D(\bar f) 
&=&
\gamma a b^{-5}c^{-5} \int\!\!\int L\,f \,dv\,dx +
b^{-3-2l}c^{-3-2l} \C(ab^{2l}c^{2l} f) \nonumber\\
&& {}
+ a b^{-3} c^{-5} \ekin(f)
+ a^2b^{-5} c^{-6}\epot (f) . \label{dscale}
\eea
To prove (a) we  
fix some $f\in \F_1$ with compact support and $L^{-l}f\leq F_0$.
Let 
\[
a=M b^3c^3
\]
so that
\[
\int\!\!\int \bar f \, dv\, dx = M.
\]
Using (Q2),
\[
\D(\bar f) \leq
\overline{C}_1 \gamma (bc)^{-2} + 
\overline{C}_2 a^{1/k_2} (bc)^{2l/k_2} + \overline{C}_3 c^{-2}
- \overline{C}_4 b  
\]
where $\overline{C}_1,\ldots,\overline{C}_4 >0$ depend on $f$,
and we need to make sure that 
$a (bc)^{2l}\leq 1$ so that (Q2) applies. Since
we want the last term to dominate
as $b \to 0$, we let $c=b^{-\eta/2}$ so that $bc = b^{1-\eta/2}$
for some $\eta >0$. Then 
\[
\D(\bar f) \leq
\overline{C}_1 \gamma b^{\eta-2} + 
\overline{C}_2 b^{(1-\eta/2)(2l+3)/k_2} + \overline{C}_3 b^\eta
- \overline{C}_4 b. 
\]
Now fix $\eta \in ]1,2[$ such that $(1-\eta/2)(2l+3)/k_2 >1$;
such an $\eta$ exists by the assumptions on $k_2$ and $l$.
For $b>0$ sufficiently small the sum of the last three
terms will be negative and $a (bc)^{2l}=M b^{(3+2l)(1-\eta/2)} <1$. 
If we fix such a $b$ then all the parameters in the above estimate are
determined in terms of $M$,
except for $\gamma$ which now can be chosen sufficiently small
to guarantee that the right hand side of the estimate
above is negative.

To show part (b) we
assume that $f \in \F_{M_2}$ and $\bar f \in \F_{M_1}$ so that by 
(\ref{mscale}),
\be \label{m1m2}
a b^{-3}c^{-3} = \frac{M_1}{M_2} =: m \leq 1.
\ee
By (\ref{dscale}) and (Q3), 
\beas
\D(\bar f) 
&=& \gamma m (bc)^{-2}\int\!\!\int L f\\
&&
{} + m\, a^{-1} (bc)^{-2l} \C(a (bc)^{2l} f) 
+ m c^{-2} \ekin(f)
+ m^2 b  \epot(f)\\
&\geq&
 \gamma m (bc)^{-2}\int\!\!\int L f
+ m a^{1/k_3} (bc)^{2l/k_3} \C(f)
+ m c^{-2} \ekin(f)
+ m^2 b  \epot(f),
\eeas
provided $a (bc)^{2l} \leq 1$. Now we require that
\[
m a^{1/k_3} (bc)^{2l/k_3}= m c^{-2} = m^2 b.
\]
Together with (\ref{m1m2}) this determines $a, b, c$ in terms of $m$.
In particular $a (bc)^{2l} =m^{2 k_3 (1+l)/(3/2 +l - k_3)} \leq 1$.  
We have
\beas
\D(\bar f) 
&\geq& 
\gamma m (bc)^{-2} \int Lf
+ m^{1+\alpha} \Bigl(\C(f) + \ekin(f) + \epot (f)\Bigr)\\
&\geq&
 m^{1+\alpha} \D(f),
\eeas
where $\alpha = (2l+2)/(l+3/2-k_3) >0$; observe that 
$m (bc)^{-2} \geq m^{1+\alpha}$ since $bc = m^{(l+1)/(l+3/2-k_3) -1}$. 
Since for any given choice of $a,\,b,\,c$ the mapping $f \mapsto \bar f$
is one-to-one and onto between $\F_{M_2}$ and $\F_{M_1}$ 
the scaling inequality follows.
\prfe  

The scaling estimate above can be used to show that along
a minimizing sequence the mass has to concentrate in a certain ball: 

\begin{lemma} \label{r0}
Let $M>0$, and let $\gamma \geq0$ be sufficiently small so that
Lemma~\ref{scaling} (a) applies.
Then there exists a radius $R_M>0$ such that
if $(f_n) \subset \F_M$ is a minimizing sequence of $\D$,
\[
\lim_{n\to \infty} \int_{|x|\ge R} \int f_n dv\,dx = 0,\ R > R_M .
\]
\end{lemma}

\noindent
{\bf Proof.} 
We define the ball 
$B_R := \{x \in \R^3 \mid |x|\le R\}$.
Let $1_{B_R\times \R^3}$ be the characteristic function of $B_R \times \R^3$.
For $f \in \F_M$ we split 
\[
f_1=1_{B_R\times \R^3} f,\ f_2 = f - f_1
\]
and let $\rho_i$ and $U_i$ denote the induced spatial densities
and potentials respectively, $i=1,2$. We abbreviate $\lambda=M-m_f(R)$.
Then
\beas
\D (f)
&=&
\D(f_1) + \D(f_2)
-\frac{1}{4\pi}\int\nabla U_1 \cdot \nabla U_2\\
&\geq&
\D_{M-\lambda} + \D_{\lambda}
-  \frac{1}{4\pi}\int\nabla U_1 \cdot \nabla U_2
\eeas
since $f_1 \in \F_{M-\lambda}$ and $f_2 \in \F_\lambda$.
Since $\nabla U_2=0$ on $B_R$,
\[
\int\nabla U_1 \cdot \nabla U_2
\leq \lambda (M-\lambda) \,
4\pi \int_R^\infty \frac{1}{r^2}dr =
\frac{4\pi}{R} \lambda (M-\lambda).
\]
Using Lemma~\ref{scaling} (b) we find that
\[
\D(f) \geq
\left[(1-\lambda/M)^{1+\alpha}+(\lambda/M)^{1+\alpha}\right] \D_M
- \frac{1}{R} \lambda (M-\lambda) .
\]
Since $\alpha>0$, there is $C_\alpha>0$, such that 
\[
(1-x)^{1+\alpha}+x^{1+\alpha}-1\le -C_\alpha (1-x)x,\ 0\le x\le 1. 
\]
Choosing $x=\lambda/M$ and noticing that by Lemma~\ref{scaling} (a)
$\D_M<0$, we have 
\bea
\D (f) - \D_M
&\geq&
\left[(1-\lambda/M)^{1+\alpha}
+(\lambda/M)^{1+\alpha}-1 \right] \D_M - 
\frac{1}{R} \lambda (M-\lambda) \nonumber\\
&\geq&
-C_\alpha \D_M 
\left(1-{\lambda \over M}\right){\lambda\over M} 
- \frac{1}{R} \lambda (M-\lambda) \nonumber\\
&=&
\left(-\frac{C_\alpha \D_M}{M^2} - \frac{1}{R}\right)\,
(M-\lambda)\lambda \nonumber \\
&=&
\left(\frac{1}{R_M} -{1\over {R}}\right)\, m_f(R)\, (M-m_f(R))
\label{split} 
\eea
where 
\[
R_M := -\frac{M^2}{C_\alpha  \D_M} >0 .
\]
Now let $(f_n) \subset \F_M$ be a minimizing sequence 
of $\D$, and assume the assertion of the lemma is wrong. Then 
there exist some $R>R_M$, 
$\lambda>0$, and a subsequence, called $(f_n)$ again, such that 
\[
\lim_{n\to \infty}
\int_{|x|\ge R}\int f_n dv\,dx = \lambda .
\]
For every $n\in \N$ we can choose $R_n>R$ such that
\[
\lambda_n := \int_{|x|\geq R_n}\int f_n dv\,dx = {1\over 2}
\int_{|x|\ge R}\int f_n dv\,dx .
\]
Then 
\[
\lim_{n\to \infty}
\int_{|x|\ge R_n}\int f_n dv\,dx =
\lim_{n\to\infty}
\lambda_n = \lambda/2>0.
\]
Applying the estimate (\ref{split}) to $B_{R_n}$ we get 
\begin{eqnarray*}
\D (f_n) - \D_M
&\geq& 
\left( \frac{1}{R_M}-\frac{1}{R_n}\right) (M-\lambda_n) \lambda_n
> \left(\frac{1}{R_M} -\frac{1}{R}\right) (M-\lambda_n) \lambda_n\\
&\to&
\left(\frac{1}{R_M} - \frac{1}{R} \right)
(M-\lambda/2) \lambda/2 > 0,\ n\to\infty,
\end{eqnarray*}
since  $0<\lambda/2<M$.
This contradicts the fact that $(f_n)$ is a minimizing 
sequence. \prfe

\section{Minimizers of $\D$}
\setcounter{equation}{0}

\begin{theorem} \label{exminim}
Let $M>0$, and let $\gamma \geq 0$ be sufficiently small
so that Lemma~\ref{scaling} (a) applies.
Let $(f_n) \subset \F_M$ be a minimizing sequence of 
$\D$. Then there is a minimizer $f_0$ and a subsequence
$(f_{n_k})$ such that 
$\D (f_0) = \D_M$, $\supp f_0 \subset B_{R_M}\times \R^3$ with 
$R_M$ as in Lemma~\ref{r0},
and $f_{n_k} \rightharpoonup f_0$ weakly in 
$L^{k_1,l} (\R^6)$.
For the induced potentials we have
$\nabla U_{n_k} \to \nabla U_0$ strongly in $L^2 (\R^3)$.
\end{theorem}

\noindent
{\bf Proof.} 
By Lemma~\ref{lower}, $(\P(f_n))$  
and thus $(f_n)$ is bounded in $L^{k_1,l} (\R^6)$,
cf.\ Lemma~\ref{rhoest}. 
Thus there exists a weakly convergent
subsequence, denoted by $(f_n)$ again:
\[
f_n \rightharpoonup f_0\ \mbox{weakly in }\ L^{k_1,l} (\R^6).
\]
Clearly, $f_0 \geq 0$ a.~e.,\ and $f_0$ is spherically symmetric.
Since by Lemma~\ref{r0}
\beas
M 
&=& 
\lim_{n \to \infty} \int_{\n{x}\leq R_1}\int_{\n{v}\leq R_2}f_n dv\,dx
+ \lim_{n \to \infty} \int_{\n{x}\leq R_1}\int_{\n{v}\geq R_2}f_n dv\,dx\\
&\leq&
\lim_{n \to \infty} \int_{\n{x}\leq R_1}\int_{\n{v}\leq R_2}f_n dv\,dx
+ \frac{C}{R_2^2}
\eeas
where $R_1 > R_M$ and $R_2 >0$ are arbitrary, it follows that
\[
\int_{\n{x}\leq R_1}\int f_0 dv\,dx = M
\]
for every $R_1 > R_M$. This proves the assertion on $\supp f_0$
and $\int\!\!\int f_0 =M$. Also by weak convergence 
\be \label{wke}
\int\!\!\int \n{v}^2 f_0 dv\,dx \leq \liminf_{n \to \infty}
\int\!\!\int \n{v}^2 f_n dv\,dx < \infty.
\ee
By Lemma~\ref{rhoest} $(\rho_n)=(\rho_{f_n})$ is bounded in 
$L^{n_1,l} (\R^3)$.
After extracting a further subsequence, we thus have
that
\[
\rho_n \rightharpoonup \rho_0:=\rho_{f_0} \
\mbox{weakly in }\ L^{n_1,l} (\R^3) .
\]
Thus by Lemma~\ref{uest} the convergence of the fields in $L^2(\R^3)$
follows.

It remains to show that $f_0$ is actually a minimizer, in
particular, $\P(f_0) < \infty$ so that $f_0 \in \F_M$.
By Mazur's Lemma there exists a sequence $(g_n) \subset L^{k_1,l}(\R^6)$ 
such that $g_n \to f_0$ strongly in $L^{k_1,l}(\R^6)$ and
$g_n$ is a convex combination of $\{f_k\mid k\geq n\}$.
In particular, $g_n \to f_0$ a.~e.\ on $\R^6$. By (Q4) the functional
\[
f \mapsto  \int\!\!\int \left(\gamma L f +  Q(L^{-l}f)L^l\right)\,dv\,dx
\]
is convex. Combining this with Fatou's Lemma implies that
\beas
\int\!\!\int \left(\gamma L f_0 +  Q(L^{-l}f_0)L^l\right)\,dv\,dx 
&\leq&
\liminf_{n \to \infty}
\int\!\!\int \left(\gamma L g_n +  Q(L^{-l}g_n)L^l\right)\,dv\,dx\\
&\leq&
\limsup_{n \to \infty}
\int\!\!\int \left(\gamma L f_n +  Q(L^{-l}f_n)L^l\right)\,dv\,dx.
\eeas
Together with (\ref{wke}) this implies that
\[
\P(f_0) \leq \lim_{n\to \infty} \P(f_n) < \infty ;
\]
note that $\lim_{n\to \infty} \P(f_n)$ exists.
Therefore,
\[
\D(f_0) = \P (f_0) - \frac{1}{8 \pi} \int \n{\nabla U_0}^2 
\leq 
\lim_{n\to \infty} \left(\P(f_n) - \frac{1}{8 \pi} \int\n{\nabla U_n}^2\right)
= \D_M,
\]
and the proof is complete. \prfe

\begin{theorem} \label{propminim}
Let $f_0 \in \F_M$ be a minimizer of $\D$. Then
\[
f_0 (x,v)=\left\{
\begin{array}{ccl} 
(Q')^{-1}(E_0-E - \gamma L) L^l &,& E_0 - E - \gamma L > 0,\\
0 &,& E_0 - E - \gamma L\le 0
\end{array}
\right.
\]
where
\[
E := \frac{1}{2} \n{v}^2 + U_0 (x),
\]
\[
E_0 := {1\over M} \int\!\!\int \left(Q'(f_0)
+ E +\gamma L \right)\,f_0\,dv\,dx  <0
\]
and $U_0$ is the potential induced by $f_0$.
Moreover, $f_0$ is a steady state of the Vlasov-Poisson
system.
\end{theorem}

\noindent
{\bf Proof.} 
Let $f_0$ be a minimizer. We shall use the standard method of 
Euler-Lagrange multipliers to prove the theorem. 
Let $\epsilon>0$, and $\eta  \in L^\infty(\R^6)$ be
compactly supported and spherically symmetric with
\[
\eta \geq 0\ \mbox{a.~e.\ on}\ \R^6 \setminus \supp f_0,\ 
\int\!\!\int \eta\,dv\,dx = 0,
\]
\[
\epsilon \leq f_0 \leq \frac{1}{\epsilon}\ 
\mbox{a.~e.\ on}\ \supp f_0 \cap \supp \eta,\
\epsilon \leq L \leq \frac{1}{\epsilon}\
\mbox{a.~e.\ on}\ \supp \eta.
\]
Below we will occasionally argue pointwise on $\R^6$ so we choose
a representative of $f_0$ satisfying the previous estimate pointwise. 
For 
\[
0 \le h \le {{\epsilon}\over{2(1+\|\eta\|_\infty)}}
\] 
we have $f_0 + h \eta \in \F_M$; that $\C(f_0 + h \eta) < \infty$
will follow from the estimates below. 
We expand $\D(f_0+h \eta)-\D(f_0)$ in powers of $h$:
\beas
\D(f_0+h\eta)-\D(f_0)
&=&
\int\!\!\int \Bigl( Q(L^{-l}(f_0+h\eta))-Q(L^{-l}f_0)\Bigr)L^l \,dv\,dx\\
&&
{}+ h  \int\!\!\int \Bigl(\gamma L + \frac{1}{2} \n{v}^2 + U_0 \Bigr)\,\eta\, dv\, dx 
 - h^2 \frac{1}{8\pi} \int \n{\nabla U_{\eta}}^2 dx ; 
\eeas
in expanding the potential energy term we used Lemma~\ref{uest} (c).
To expand the first term
we first consider a point $(x,v) \in \supp f_0 \cap \supp \eta$. Then
\[
\Bigl(Q(L^{-l}(f_0+h\eta))-Q(L^{-l}f_0)\Bigr) L^l
=
h Q'(L^{-l}f_0) \eta + 
h^2 \frac{1}{2} Q''(\tau)L^{-l} \eta^2
\]
where $\tau$ lies between $L^{-l} f_0$ and $L^{-l}(f_0+h\eta)$. 
Thus
\[
\Bigl|\Bigl(Q(L^{-l}(f_0+h\eta))-Q(L^{-l}f_0)\Bigr) L^l
 - h Q'(L^{-l}f_0) \eta \Bigr| 
\leq  C h^2 \eta^2;
\]
note that $0 < c_1 \leq L^{-l} f_0,\, L^{-l} (f_0 + h \eta) \leq c_2 $
on $\supp \eta \cap \supp f_0$ for constants $c_1,\, c_2 > 0$,
and $Q''$ is continuous on the interval $[c_1,c_2]$. 
On $\supp \eta \setminus \supp f_0$  the assumption $(Q2)$ implies
that for $h$ small 
\beas
\Bigl|\Bigl(Q(L^{-l}(f_0 + h \eta))-Q(L^{-l}f_0)\Bigr) L^l
 - h Q'(L^{-l}f_0) \eta\Bigr| 
&=&  Q(L^{-l} h \eta) L^l\\
&\leq&
C \n{\eta}^{1+1/k_2} h^{1+1/k_2}.
\eeas
The fact that $f_0$ is a minimizer 
and the estimates above imply that
\beas
0 \leq \D(f_0 + h \eta) - \D(f_0)
&=&
h \int\!\!\int  
\left(Q'(L^{-l}f_0) + \gamma L + \frac{1}{2} \n{v}^2 + U_0 \right)
\,\eta \, dv\,dx\\
&& 
{} + O(h^{1+\delta})
\eeas
for all $h>0$ sufficiently small. Recalling the definition of $E$
this implies 
\[
\int\!\!\int \Bigl(Q'(L^{-l} f_0)
+ E +\gamma L \Bigr)\, \eta\, dv\, dx \geq 0
\]
for all admissible $\eta$, with equality if 
$\supp \eta \subset\supp f_0$ 
since then also $-\eta$ is admissible. Recalling the
definition of $E_0$ we obtain
\beas
\int\!\!\int \Bigl(Q'(L^{-l} f_0)
+ E +\gamma L \Bigr)\, (f_0 + \eta)\, dv\, dx 
&\geq& \int\!\!\int \Bigl(Q'(L^{-l} f_0)
+ E +\gamma L \Bigr)\, f_0\, dv\, dx \\
&=&
E_0 M =
E_0 \int\!\!\int (f_0 + \eta)\, dv\,dx,
\eeas
or
\[
\int\!\!\int \Bigl(Q'(L^{-l} f_0)
+ E +\gamma L - E_0 \Bigr)\, (f_0 + \eta) \, dv\, dx \geq 0,
\]
again with equality if $\supp \eta \subset \supp f_0$.  
Recalling the class of admissible test functions $\eta$ 
and the fact that $\epsilon >0$ is arbitrary we conclude that 
\[
Q'(L^{-l} f_0) +  E + \gamma L - E_0 = 0 \ \ \mbox{a.~e.\ on}\
\supp f_0,
\]
and
\[
E +\gamma L - E_0 \geq 0 \ \ \mbox{a.~e.\ on}\ \R^6 \setminus \supp f_0 .
\]
To see the former define 
$g:=Q'(L^{-l} f_0) + E +\gamma L - E_0$ and take spherically symmetric,
measurable sets $B^+, B^- \subset \supp f_0$ such that
$g>0$ on $B^+$, $g<0$ on $B^-$, and $g=0$ on
$\supp f_0 \setminus(B^+ \cup B^-)$. For $\epsilon>0$ define
\[
K_\epsilon := 
\{ (x,v) \in \R^6 | \epsilon \leq f_0(x,v), L(x,v) \leq 1/\epsilon \};
\]
here $g$ and $f_0$ are understood as representatives of the corresponding
a.~e.\ equivalence classes of measurable functions. 
Define $B_\epsilon^\pm := B^\pm \cap K_\epsilon$, and assume
that ${\rm vol} B^+ >0$ and thus also ${\rm vol} B^+_\epsilon >0$
for $\epsilon>0$ sufficiently small. Now define
$\eta := \alpha f_0$ on $B^+_\epsilon$,
$\eta := - f_0$ on $B^-_\epsilon$ and zero elsewhere, where
$\alpha \geq 0$ is such that $\int \eta =0$; note that
for $\epsilon> 0$ sufficiently small, $\int_{B^+_\epsilon} f_0 >0$.
This $\eta$ is admissable, in particular it has support in the 
set $K_\epsilon \subset \supp f_0$. Thus
\[
0 = \int g\,(f_0+\eta) = \int_{B^+_\epsilon} g \,(1+\alpha)\, f_0
+ \int_{\R^6\setminus K_\epsilon} g\, f_0 \geq \int_{B^+_\epsilon} g\, f_0
+ \int_{\R^6\setminus K_\epsilon} g\, f_0 
\]
where the first integral is positive and increasing with $\epsilon \to 0$
and the second converges to 0. This is a contradiction so that 
${\rm vol} B^+ = 0$. The same argument works for $B^-$ so that
$g=0$ on $\supp f_0$. Thus 
$0 \leq \int_{\R^6\setminus \suppi f_0} g \eta$ for all admissable $\eta$
which implies that $g \geq 0$ outside $\supp f_0$. 

This implies that $f_0$ is of the form given in the theorem,
and by construction
\[
\lap U_0 = \frac{1}{r^2} (r^2 U_0')' = 4 \pi \rho_0
\]
so that $(f_0,\rho_0,U_0)$ is indeed a solution of the Vlasov-Poisson
system. Since $f_0$ has compact support and 
$\lim_{r \to \infty} U_0 (r)=0$ we conclude that $E_0<0$. \prfe

\section{Dynamical stability}
\setcounter{equation}{0}

We now investigate the dynamical stability of $f_0$. First we note
that for $f \in \F_M$,
\begin{equation}
\D (f)- \D (f_0)=d(f,f_0)-\frac{1}{8 \pi}
\|\nabla U_f-\nabla U_0\|^2_2.\label{d-d}
\end{equation}
where
\[
d(f,f_0) = \int\!\!\int \Bigl[Q(L^{-l}f)L^l-Q(L^{-l}f_0)L^l +
(E + \gamma L - E_0)(f-f_0)\Bigr]\,dv\,dx.
\]

\begin{theorem} \label{stability}
Let $Q$ satisfy the assumptions (Q1)--(Q4) and
assume that the minimizer $f_0$ is unique in $\F_M$.
Then for all $\epsilon>0$ there is $\delta>0$ such that
for any solution $f(t)$ of the Vlasov-Poisson system
with $f(0) \in C^1_c (\R^6)\cap \F_M$,
\[
d(f(0),f_0) + \frac{1}{8\pi} \|\nabla U_{f(0)}-\nabla U_0\|_2^2 < \delta
\]
implies
\[
d(f(t),f_0) + \frac{1}{8\pi} \|\nabla U_{f(t)}-\nabla U_0\|_2^2 < \epsilon,
\ t \geq 0.
\]
\end{theorem}

\noindent
{\bf Proof.}
We first show that $d(f,f_0) \geq 0,\ f \in \F_M$.
For $E + \gamma L - E_0 \geq 0$ we have $f_0 = 0$, and thus
\[
Q(L^{-l}f)L^l-Q(L^{-l}f_0)L^l + (E + \gamma L - E_0)(f-f_0) 
\geq Q(L^{-l}f)L^l \geq 0.
\]
For $E +\gamma L - E_0 <  0$,
\beas
Q(L^{-l}f)L^l - Q(L^{-l}f_0)L^l + (E + \gamma L - E_0)(f-f_0) 
&=& 
\frac{1}{2} Q''(L^{-l}\tilde f)L^{-l} (f - f_0)^2\\
&\geq& 0
\eeas
provided $f>0$; here $\tilde f$ is between $f$ and $f_0$. If $f=0$,
the left hand side is still nonnegative by continuity.

We will use the fact that $\D$ is conserved along
any solution $f(t)$ of the Vlasov-Poisson system with
$f(0)\in C^1_c (\R^6)\cap \F_M$, i.~e., along any classical, spherically
symmetric solution. This follows from conservation of energy and
the fact that
both $f(t)$ and $L$ are constant along the measure preserving  
characteristic flow.
Assume the assertion of the theorem were false. 
Then there exist $\epsilon_0>0$, $t_n>0$, and
$f_n(0) \in C^1_c (\R^6)\cap \F_M$ such that 
\[
d(f_n(0),f_0) + 
\frac{1}{8\pi} \|\nabla U_{f_n(0)}-\nabla U_0\|_2^2 = \frac{1}{n}
\]
but
\[
d(f_n(t_n),f_0) +
\frac{1}{8\pi} \|\nabla U_{f_n(t_n)}-\nabla U_0\|_2^2 \ge \epsilon_0>0.
\]
From (\ref{d-d}), we have 
$\lim_{n\to\infty} \D (f_n(0))=\D_M$. 
Since  $\D (f)$ is invariant under the Vlasov-Poisson flow,
\[
\lim_{n\to\infty} 
\D (f_n(t_n))=\lim_{n\to\infty} 
\D (f_n(0))=\D_M.
\]
Thus, $(f_n(t_n)) \subset \F_M$
is a minimizing sequence of $\D$, and by Theorem~\ref{exminim} ,
we deduce that---up to a 
subsequence---$\|\nabla U_{{f_n}(t_n)}-\nabla U_0\|^2_2\to 0$. 
Again by (\ref{d-d}), 
$d(f_n(t_n),f_0)\to 0$, a contradiction. \prfe

If $Q''$ allows an appropriate bound from below we can obtain an estimate
for a weighted $L^2$-norm of $f(t)-f_0$.  
If the minimizer $f_0$ of $\D$ is not unique in
$\F_M$ we denote by $\M_M$ the set of all minimizers of 
$\D$ in $\F_M$. Then for each $\epsilon >0$ there exists $\delta >0$
such that for any solution $f(t)$ of the Vlasov-Poisson
system with $f(0) \in \F_M \cap C^1_c(\R^6)$,
\[
\inf_{f_0 \in \M_M} 
\left[d(f(0),f_0) + \frac{1}{8\pi} \|\nabla U_{f(0)}-\nabla U_0\|_2^2\right]
 < \delta
\]
implies
\[
\inf_{f_0 \in \M_M}
\left[d(f(t),f_0) + 
\frac{1}{8\pi} \|\nabla U_{f(t)}-\nabla U_0\|_2^2\right] < \epsilon,
\ t \geq 0.
\]
The proof works along the same lines as for Theorem~\ref{stability}.

\bigskip

{\bf Final Remarks:}
\begin{itemize}
\item[(a)]
The uniqueness of the minimizer $f_0$ can be shown in the case of the 
polytropic ansatz (\ref{poly}). In the general case we know of no such result,
but we mention that for the argument in Theorem~\ref{stability}
it would suffice if the minimizers of $\D$ in $\F_M$ were isolated.
\item[(b)]
Obviously we obtain stability only against
spherically symmetric perturbations. One reason is that the quantity
$L$ is conserved by the characteristic flow only for spherically symmetric
solutions. The other is that the splitting estimate (\ref{split})
in Lemma~\ref{r0} relied on the symmetry which therefore is required
even if there is no dependence on $L$.
\item[(c)]
As was pointed out in the introduction, the polytropic ansatz
(\ref{poly}) leads to steady states with finite mass and compact
support for $k < 3 l + 7/2$. Using a scaling argument as in
the proof of Lemma~\ref{scaling} it can be shown that 
$\D_M = -\infty$ for $k>l+3/2$, cf.\ \cite{G},
so that our method does not work for
this parameter range. The same is true for the methods used
in \cite{Wo} as well as in \cite{RR}. Thus $k=l+3/2$ seems to be some kind of
threshold for the stability properties of steady states. On the other hand
it is shown in \cite{A}
that the so called Plummer's sphere obtained
for $k=3 l +7/2$ is the unique minimizer of the total
energy of the system under a more restrictive constraint;
this model has finite mass but is supported on the whole space $\R^3$.
\end{itemize}

\end{document}